\documentclass[10pt]{iopart}
\usepackage{iopams}

\usepackage[]{graphicx}
\usepackage{url}
\usepackage[T1]{fontenc}
\usepackage{dcolumn}
\usepackage{sistyle}
\usepackage{color}

\begin{document}

\title{Realization of the farad from the dc quantum Hall effect with digitally-assisted impedance bridges}
\author{Luca Callegaro\dag\   
\footnote[3]{Corresponding author (l.callegaro@inrim.it)}, Vincenzo D'Elia\dag, and Bruno Trinchera\dag
}

\address{\dag\ Istituto Nazionale di Ricerca Metrologica (INRIM), Str.\ delle Cacce 91, 10135 Torino, Italy}

\begin{abstract}

A new traceability chain for the derivation of the farad from dc quantum Hall effect has been implemented at INRIM. Main components of the chain are two new coaxial transformer bridges: a resistance ratio bridge, and a quadrature bridge, both operating at \SI{1541}{Hz}. The bridges are energized and controlled with a polyphase direct-digital-synthesizer, which permits to achieve both main and auxiliary equilibria in an automated way; the bridges and do not include any variable inductive divider or variable impedance box. The relative uncertainty in the realization of the farad, at the level of \SI{1000}{pF}, is estimated to be \SI{64E-9}{}. A first verification of the realization is given by a comparison with the maintained national capacitance standard, where an agreement between measurements within their relative combined uncertainty of \SI{420E-9}{} is obtained.
\end{abstract}



\maketitle

\section{Introduction}

A number of National Metrology Institutes work on measurement setups to trace the farad to the representation of the ohm given by the dc quantum Hall effect \cite{Jeffery98, Nakamura99, Chua99, Hsu00, Bohacek01, Small01, NPL01, Awan03, Melcher03, Inglis03, Schurr09} . The traceability chains employed involve a number of experimental steps, usually requiring not less than three different coaxial ac bridges. The bridges are complex networks of passive electromagnetic devices \cite{Kibble84}: some of such devices are fixed (transformers and single-decade inductive voltage dividers), and several are variable (resistance and capacitance boxes, multi-decadic inductive voltage dividers). Bridges are balanced by operating on the variable devices to reach equilibrium; that is, the detection of zero voltage or current on a number of nodes in their electrical networks. Variable devices, especially inductive voltage dividers, are typically manually operated; only a few models have been described \cite{Ramm85,Avramov93,NPL01,Giblin07} \footnote{A commercial item is the Tegam mod. PRT-73.} which permit remote control. Consequently, a large part of existing bridges are manually operated.

In most cases, the role of variable devices in a bridge is to synthesize signals (voltages or currents) to be injected in the bridge network to bring a detector position to zero. The signals are isofrequential with the main bridge supply, and can be adjusted in their amplitude and phase relationships. The amplitude and phase of one (or more) of the signals enters the measurement model equation, and must be calibrated (\emph{main balance}), but the others (\emph{auxiliary balances}) don't need a calibration.

In this view, it is straightforward to consider a substitution in the bridge network of most, or all, variable passive devices with a corresponding number of \emph{active} sinewave generators, locked to the same frequency but adjustable in amplitude and phase independently of each other. Direct digital synthesis (DDS)  of sinewaves  \cite{DDS99} is a well-established technique that permit the realization of such generators; hence, in this sense, we may speak of \emph{digitally-assisted impedance bridges} when DDS generators are used. Digitally-assisted bridges impedance have been considered both theoretically  \cite{Helbach87,Tarach93} and in a number of implementations \cite{Bachmair80, Helbach83, Cabiati85, Ramm85AC, Cabiati87, Waltrip95, Muciek97, Dutta01, CallegaDDS01, Callegaro02, Corney03}; some commercial impedance meters are also digitally-assisted.

In this paper, we consider for the first time the feasibility of a complete ohm to farad traceability chain based on digitally-assisted bridges. We constructed a resistance ratio bridge and a quadrature bridge, to measure capacitance (at the level of \SI{1000}{pF}) in terms of dc quantum Hall resistance (at the level of $R_\mathrm{K}/2 \approx $\SI{12906.4}{\ohm}, where $R_\mathrm{K}$ is the Von Klitzing constant). The bridges are automated, and a single measurement can be conducted in  minutes. The estimated relative uncertainty of the capacitance determination related to the traceability chain is \SI{64E-9}{}. 

This accuracy claim hasn't yet been verified with a comparison with other farad realization. However, by completing the chain with an older manual transformer ratio bridge \cite{Cabiati78}, we performed a comparison between the new realization and the present national capacitance standard, maintained as a group value at the level of \SI{10}{pF} with a relative uncertainty of \SI{400E-9}. The measurement results of the comparison are compatible within the relative compound uncertainty.

\section{The traceability chain}
\begin{figure}[ht]
    \centering
    \includegraphics[width=2.5 in]{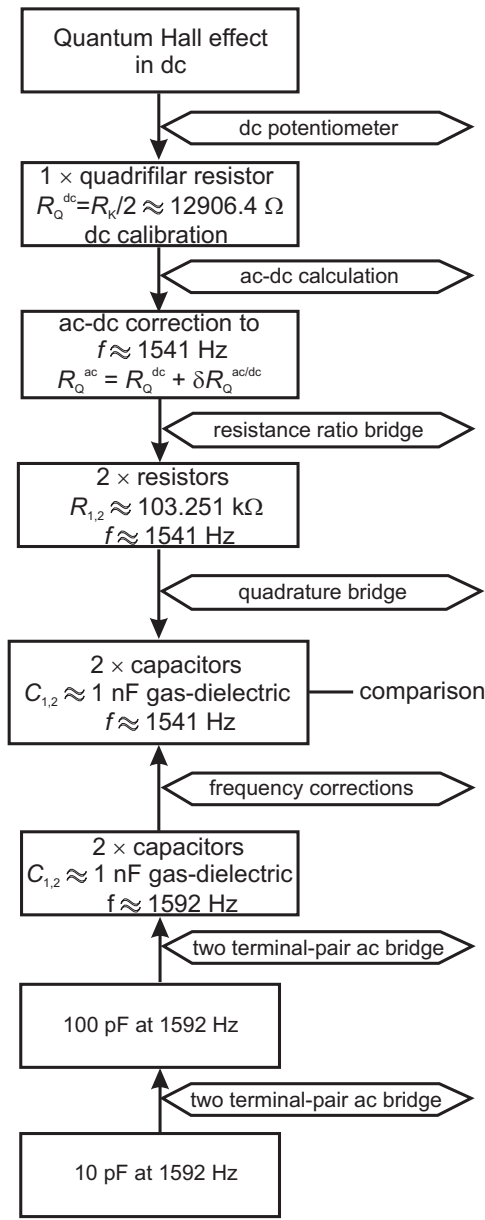}
    \caption{Graphical representation of the traceability chain for the realization of the farad unit from the quantum Hall effect.}
    \label{fig:TraceabilityChain}
\end{figure}

The traceability chain which has been set up, including the steps for the comparison (Sec. \ref{sec:comparison}) with the maintained capacitance standard is shown in graphical form in Fig. \ref{fig:TraceabilityChain}. Its steps are here summarized and will be described in more detail in the course of the paper.

\begin{itemize}
\item The quantum Hall effect is employed to calibrate a resistor $R_\mathrm{Q}$ having the nominal value $R_\mathrm{K}/2 \approx$ \SI{12906.4}{\ohm} and a calculable frequency performance;
\item $R_\mathrm{Q}$ is employed in a 8:1 resistance ratio bridge to calibrate two resistance standards $R_{1,2}$, with nominal value $4 \times R_\mathrm{K} \approx$ \SI{103.251}{k\ohm}, at the frequency $f\approx$\SI{1541.4}{Hz};
\item $R_{1,2}$ are employed in a quadrature bridge to calibrate the product $C_1 \times C_2$ of two capacitance standards $C_{1,2}$ with nominal value of \SI{1000}{pF};
\item in order to perform the comparison with the maintained capacitance standard, maintained at the level of \SI{10}{pF}, a capacitance ratio bridge is employed to perform a scaling up to \SI{1000}{pF} and the measurement of $C_1$ and $C_2$;
\item a small calculated frequency correction is applied to permit the comparison.
\end{itemize}

\section{Impedance standards}
\label{sec:ImpedanceStandards}

\begin{figure}[ht]
    \centering
    \includegraphics[width=3.5 in]{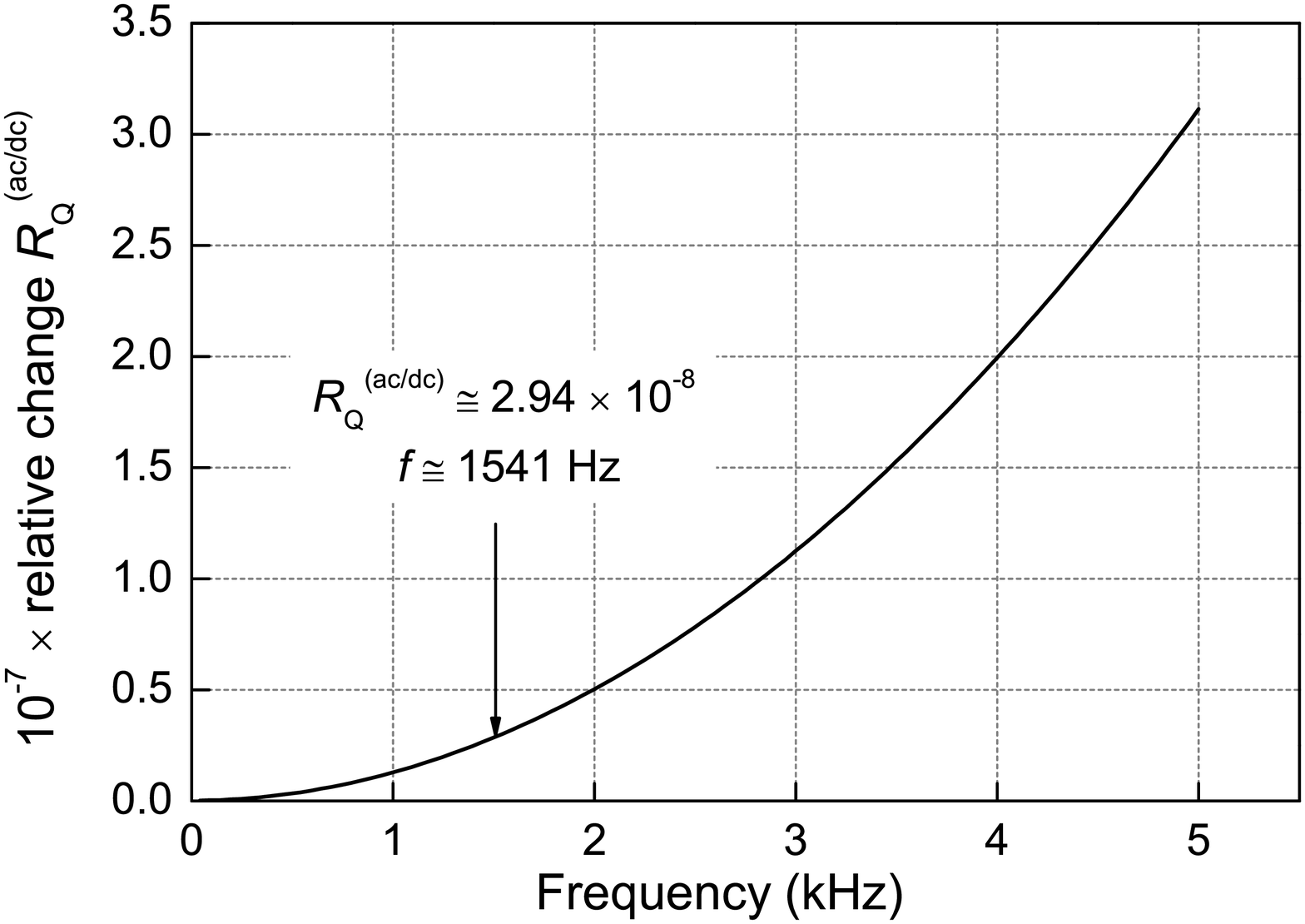}
    \caption{Frequency performance of the $R_\mathrm{Q}$ resistor.}
    \label{fig:QuadrifilarFrequencyPerformance}
\end{figure}

The standards employed in the traceability chain are:

\begin{description}

\item[$R_\mathrm{Q}$] a quadrifilar resistance standard, having a nominal value $R_\mathrm{Q} = R_\mathrm{K}/2$.\footnote{NL engineering Type QF, serial 1294.} Its frequency performance and reactive parameter are calculable from geometrical dimensions \cite{Gibbings63}: the result of the calculation is shown in Fig. \ref{fig:QuadrifilarFrequencyPerformance}. The standard is thermostated to improve its stability. The original standard has an inductance of $\approx$\SI{3}{\micro H}; in order to reduce its phase angle, a small gas-dielectric capacitor has been added in parallel to its current terminals.  

\item[$R_{1,2}$] two resistance standards with nominal value $R_{1,2} = 4 \times R_\mathrm{K}$. Presently, two thin film resistors\footnote{Vishay mod. VHA512 bulk metal foil precision resistors, \SI{\pm 0.001}{\%} tolerance, \SI{0.6}{ppm/\degC} temperature coefficient.} encased in a metal shield and defined as two terminal-pair standards are employed. The casings are within a single air bath, having \SI{1}{mK} temperature stability. Two new standards with independent thermostats are under construction.

\item[$C_{1,2}$] two gas-dielectric capacitance standards $C_{1,2} =$\SI{1}{nF} are constructed from General Radio 1404-A standards, re-encased in a thermostated bath at \SI{23}{\degC} with \SI{1}{mK} stability and redefined as two terminal-pair impedance standards. Ref. \cite{NPL01} gives a detailed description of the construction and characterization.

\end{description}

\section{Digitally-assisted coaxial bridges}

\begin{figure}[ht]
    \centering
    \includegraphics[width=3.5 in]{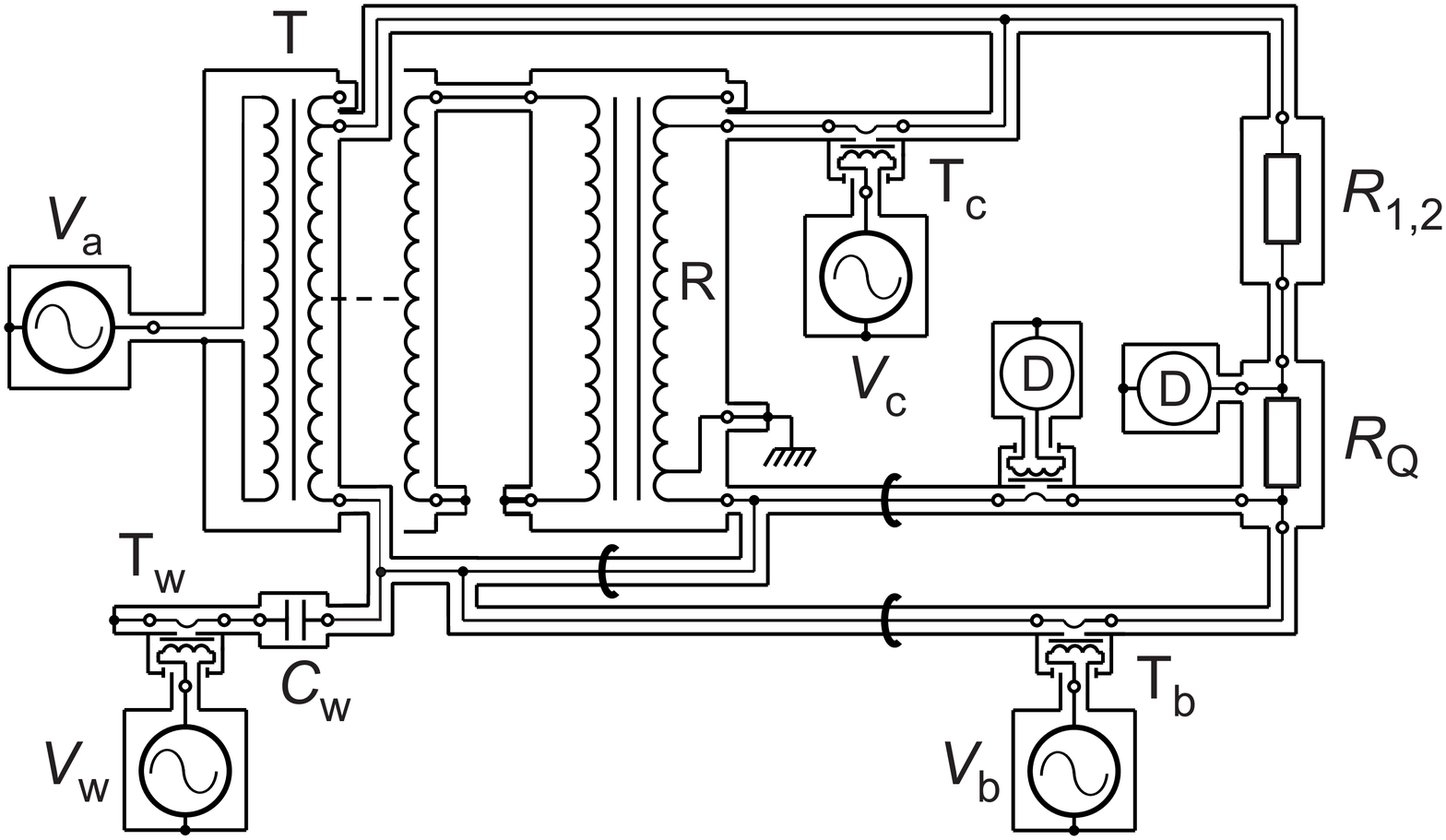}
    \caption{Simplified diagram of the 8:1 resistance ratio bridge. Black rings along the mesh are current equalizers.}
    \label{fig:RatioBridgeSchematics}
\end{figure}

\begin{figure}[ht]
    \centering
    \includegraphics[width=3.5 in]{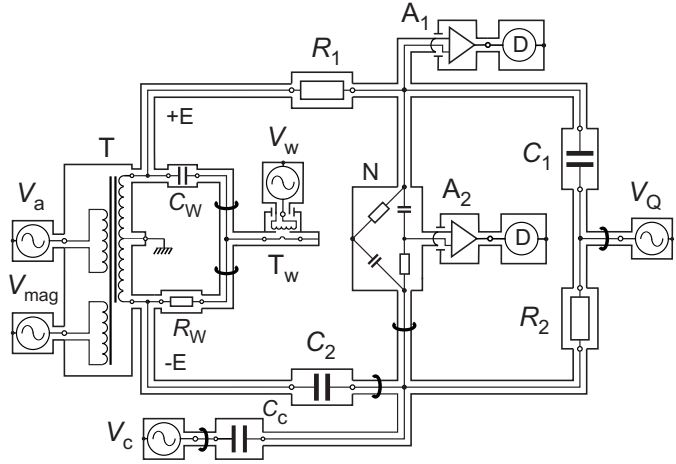}
    \caption{Simplified diagram of the two-terminal-pair quadrature bridge.}
    \label{fig:QuadBridgeSchematics}
\end{figure}

\begin{figure}[ht]
    \centering
    \includegraphics[width=3.0 in]{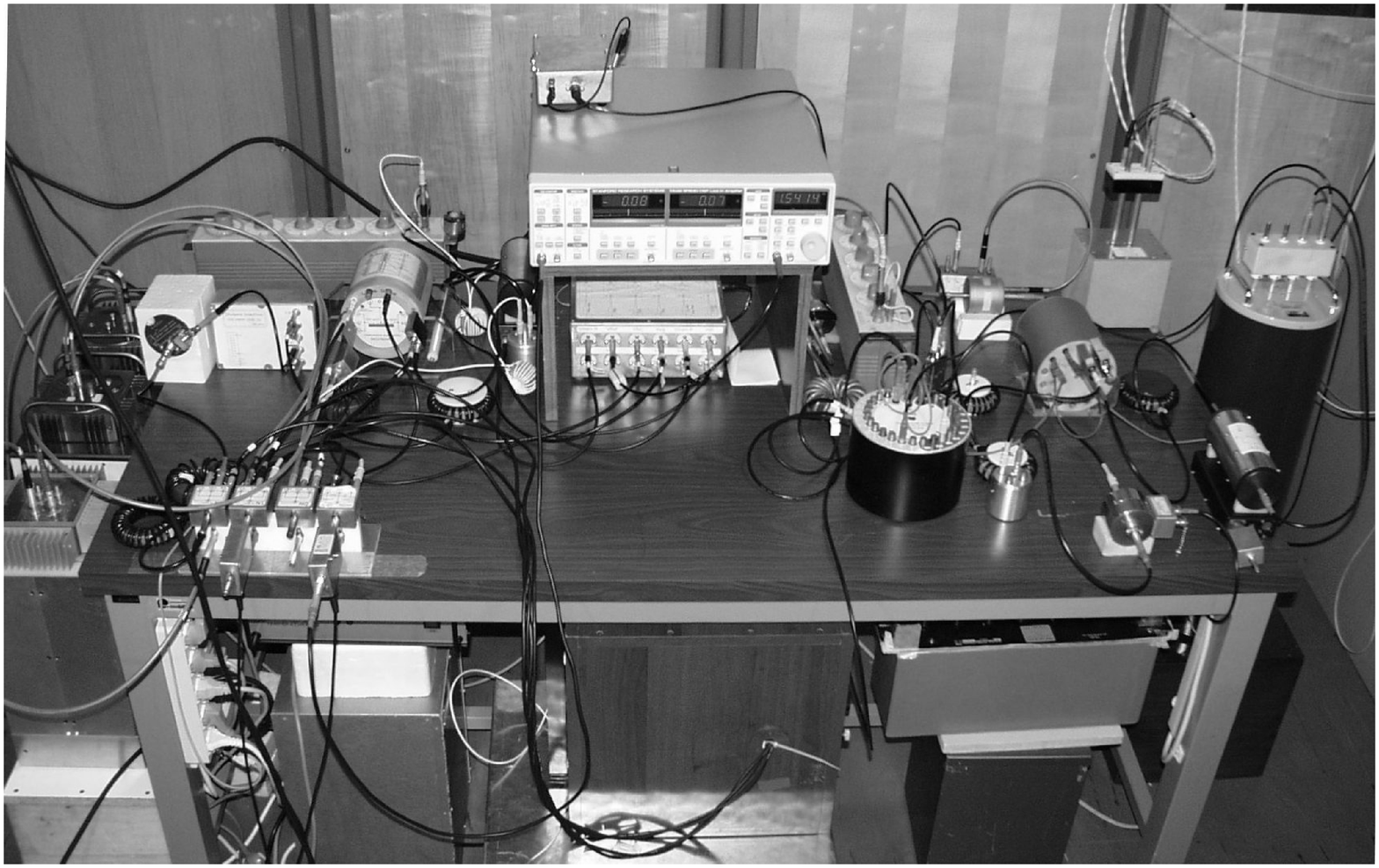}
    \caption{A photo of the two bridges. On the left the quadrature bridge; on the right the resistance ratio bridge. The instruments in the middle are common to both bridges.}
    \label{fig:Photo}
\end{figure}

The digitally-assisted bridges developed are a 8:1 resistance ratio bridge, and a quadrature bridge. The coaxial schematics can be seen in Fig. \ref{fig:RatioBridgeSchematics} and Fig. \ref{fig:QuadBridgeSchematics} respectively. The bridges are based on the same design concept and share common instrumentation: the polyphase generator (Sec. \ref{sec:PolyphaseGenerator}), the impedance standards (Sec. \ref{sec:ImpedanceStandards}), and the detector. A photo of both bridges is shown in Fig. \ref{fig:Photo}.

\subsection{Polyphase generator}
\label{sec:PolyphaseGenerator}

\begin{figure}[ht]
    \centering
    \includegraphics[width=3.5 in]{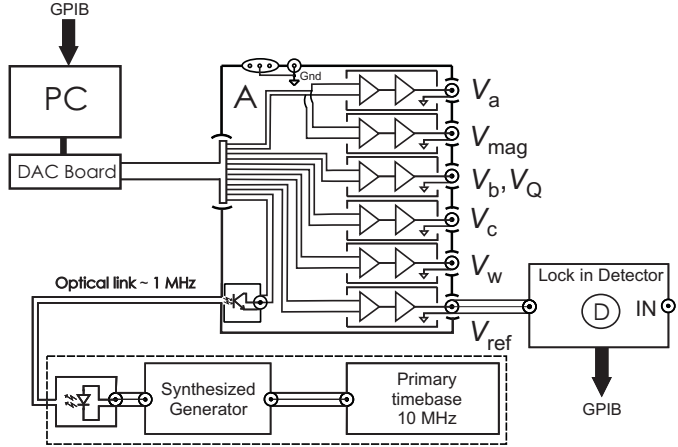}
    \caption{Schematic diagram of the sinewave synthesizer, see Sec. \ref{sec:PolyphaseGenerator} for details.}
    \label{fig:Generator}
\end{figure}

Both bridges are energized (one at a time) by a polyphase DDS generator; the schematic diagram is reported in Fig. \ref{fig:Generator}. The core of the generator is a commercial digital-to-analogue (DAC) board\footnote{National Instruments mod. NiDaq-6733 PCI board, 8 DAC outputs, variable reference input, 16 bit resolution, maximum sampling rate \SI{1}{MS\, s^{-1}}, voltage span \SI{\pm 10}{V}.}. The board is programmed for a continuous generation of sinewaves; each wave can be updated without stopping the generation (large amplitude or phase changes are gradually achieved to avoid steps in the output).

Since the sinewaves are represented by an integer number of samples (presently 628), the output frequency is finely tuned by changing the common DAC update clock frequency, typically ranging between \SI{950}{kHz} and \SI{1}{MHz}. The clock is given by a commercial synthesizer\footnote{Stanford Research System mod. DS345.} connected to the DAC board by an optical fibre link \cite{Callegaro99} to minimize high-frequency interferences. The synthesizer is in turn locked to INRIM \SI{10}{MHz} timebase; hence, the frequency uncertainty of the polyphase generator is better than \SI{1E-10}{}.

Five DAC channels are used. Four are employed on the bridge network, the fifth gives a reference signal for the lock-in amplifier which acts as zero detector. The four channels enter a purposely-built analog electronics which include, for each channel, a line receiver (which decouple the computer ground from the bridge ground), a \SI{200}{kHz} low pass two-pole Butterworth filter  to reduce the quantization noise, and a buffer amplifier with automatic control of dc offset \cite{BurrBrown90} to avoid possible magnetizations of the electromagnetic components. The analog gain of each channel can be finely trimmed.   

\subsection{Resistance ratio bridge}
\label{sec:RatioBridge}

A simplified coaxial diagram is shown in Fig. \ref{fig:RatioBridgeSchematics}. Output $V_\mathrm{a}$ of the polyphase generator energizes the main isolation transformer T, which has two secondary windings: one supplies the measurement current, the other energizes the magnetizing winding of the main ratio divider R. $R_Q$ is defined as a four terminal-pair impedance, whereas $R_{1,2}$ is defined as two terminal-pair impedance. Output $V_\mathrm{b}$ of the generator, with injection transformer T$_\mathrm{b}$, adjusts the current in $R_\mathrm{Q}$.
Output $V_\mathrm{c}$, and injection transformer T$_\mathrm{c}$ having ratio $D_\mathrm{c}$, provide the main balance by adjusting R ratio. Output $V_\mathrm{W}$, with transformer T$_\mathrm{W}$ and injection capacitance $C_\mathrm{W}$, provides Wagner balance. The detector D is the input of the lock-in amplifier (in floating mode), manually switched between detection points.

The result of measurement $R_{1,2}/R_\mathrm{Q} \approx 8$ can be expressed as 
\begin{equation}
\label{eq:ratiobridgeequilibrium}
\frac{R_{1,2}}{R_\mathrm{Q}} = 8 \left\{ 1 + \frac{10}{8} \cdot \frac{1}{D_\mathrm{c}} \cdot \frac{\left| V_\mathrm{c} \right|}{\left| V_\mathrm{a} \right|} \cos \left[ \arg(V_\mathrm{c}) - \arg(V_\mathrm{a}) \right] - \frac{81}{8} \epsilon_\mathrm{ph}  \right\},
\end{equation}

Eq. \ref{eq:ratiobridgeequilibrium} takes into account also the complex deviation $\epsilon$ of the ratio $k$ of R from its nominal value $1/9$, expressed as $k = 1/9 + \epsilon_\mathrm{ph} + j \epsilon_\mathrm{qd}$. R is calibrated using a bootstrap technique \cite{Callegaro2003IVD} validated in an international intercomparison \cite{CCEMK7}.

\subsection{Quadrature bridge}
\label{sec:QuadBridge}

The quadrature bridge is shown in Fig. \ref{fig:QuadBridgeSchematics}. The bridge measure the product $C_1 C_2$ in terms of $R_1 R_2$; it is an evolution of a similar bridge presented in \cite{Trinchera09}.

The main ratio transformer T has a magnetizing winding (driven by generator output $V_\mathrm{mag}$) and a primary winding (driven by output $V_\mathrm{a}$). The secondary winding is a center-tapped bifilar winding providing two nominally equal outputs +E and -E. 

The double equilibrium of the quadrature bridge is obtained by adjustments of the quadrature voltage (provided by generator output $V_Q$) and of a balancing current (provided by output $V_\mathrm{c}$ and an injection capacitor $C_\mathrm{c}$). 

A fixed combining network N decouples the adjustments; detector points are monitored with low-noise amplifiers $\mathrm{A_1}$ and $\mathrm{A_2}$ and the lock-in amplifier, manually switched between the two detection points \footnote{The notch filter for harmonics rejection, commonly employed in other setups \cite{Kibble84, Schurr09}, has proven unnecessary because of the high harmonic rejection (\SI{-90}{dB}) of the digital lock-in amplifier employed, Stanford Research Systems model SR830, and of $\mathrm{A_1}$ and $\mathrm{A_2}$. The residual effect has been considered as an uncertainty contribution, see Sec. \ref{sec:uncertainty}.}.

Output $V_\mathrm{W}$, with transformer T$_\mathrm{W}$ and injection network $C_\mathrm{W}$-$R_\mathrm{W}$, provides Wagner balance.

The reading of the quadrature bridge can be expressed as (see Ref. \cite{Kibble84}, ch. 6.2.2)

\begin{equation}
\label{quadbridgeequilibrium}
\omega^2 R_1 R_2 C_1 C_2 = 1 + \delta
\end{equation}

The real part $\mathrm{Re}\left[ \delta \right]$ of $\delta$, which links principal values of $R_1, R_2, C_1, C_2$ (the imaginary part $\mathrm{Im} \left[ \delta \right]$ links resistance time constants to capacitance losses) is given by the expression 
 
\begin{equation}
\label{eq:deltavalue}
\mathrm{Re} \left[ \delta \right] = \frac{\left| V_\mathrm{c} \right|}{\left| V_\mathrm{Q} \right|} \, \omega \, C_\mathrm{c} \, R_2 \sin \left[ \arg(V_\mathrm{c}) - \arg(V_\mathrm{Q}) \right]
\end{equation}

where $\left| V_\mathrm{c} \right|$ and $\left| V_\mathrm{Q} \right|$ are the amplitudes of phasors $V_\mathrm{c}$ and $V_\mathrm{Q}$, and $\arg(V_\mathrm{c})$, $\arg(V_\mathrm{Q})$ are their phases.

Eq. \ref{eq:deltavalue} does not take into account possible asymmetries of transformer T; however, these are compensated by exchanging the connections of the outputs of T to the bridge network, and by averaging the two values of $\mathrm{Re} \left[ \delta \right]$ obtained with the two equilibria.

\subsection{Bridge operation}

The bridges are operated in a similar way, with the same control program. The user interface permits to set the amplitude and relative phase of each generator output; to achieve equilibrium, an automated procedure \cite{Callegaro05} is implemented, resulting an increased speed and ease of operation. Presently the detector input must be manually switched between different detection points; despite this, equilibrium is reached from an arbitrary setting in a few minutes; if the if the bridge is already near equilibrium condition the procedure is faster.

\section{Maintained capacitance unit}

The Italian capacitance national standard is presently maintained as the group value of several \SI{10}{pF} quartz-dielectric capacitors \cite{Cabiati78}. The capacitance differences are periodically monitored, and the group value is updated by drift prediction and by participating to international comparisons \cite{Belliss02}. The scaling from \SI{10}{pF} to \SI{1000}{pF} is performed with a manual two terminal-pair coaxial ratio bridge \cite{Cabiati78} and a step-up procedure which permits to compensate for possible deviations of the transformer ratio from its nominal value.

\section{Results}
\label{sec:results}

\subsection{Measurements of $R_\mathrm{Q}$}
\label{sec:dcmeasurements}

\begin{figure}[ht]
    \centering
    \includegraphics[width=3.5 in]{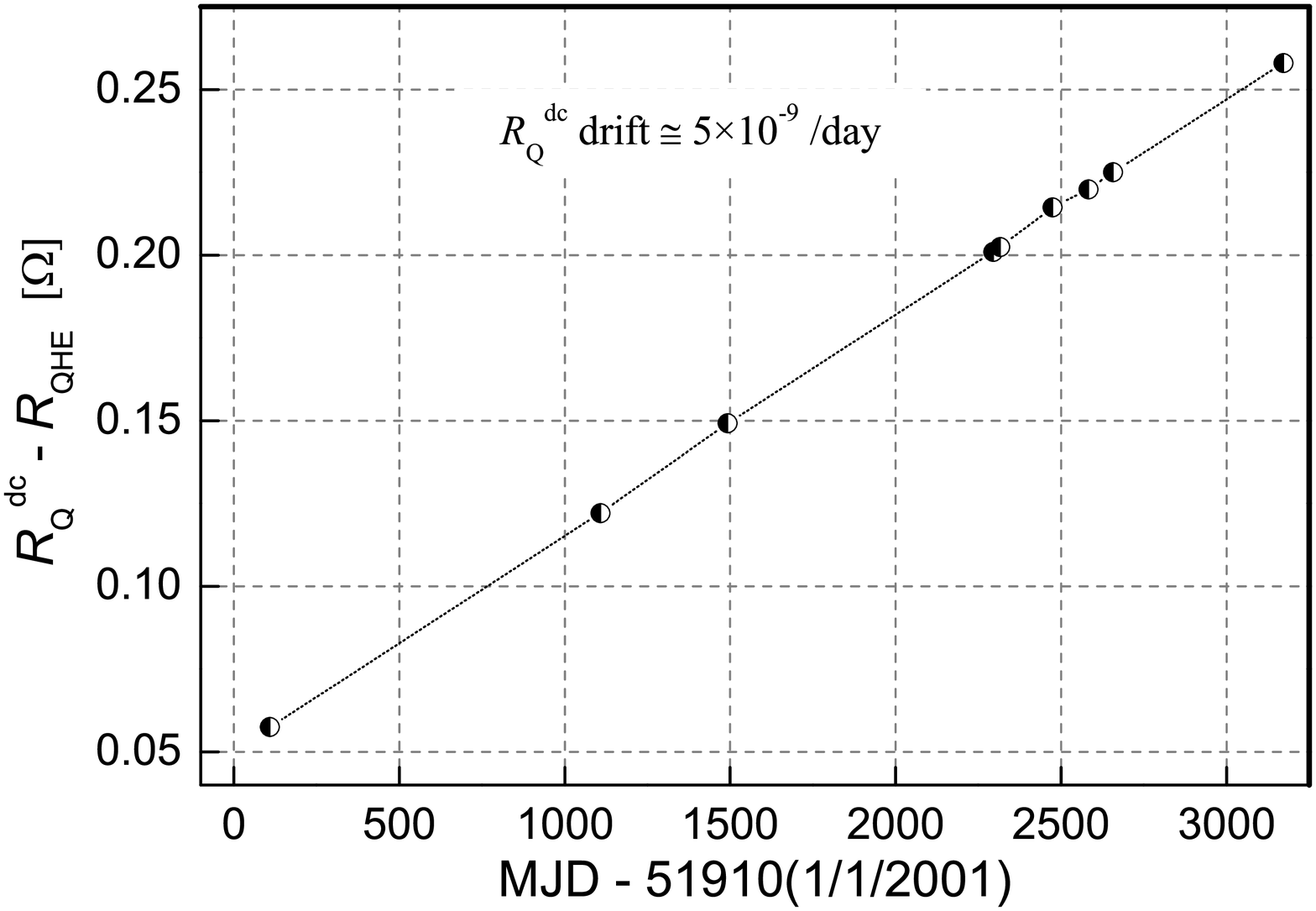}
    \caption{Time drift of the quadrifilar resistor $R_\mathrm{Q}$.}
    \label{fig:QuadrifilarDrift}
\end{figure}

The representation of the ohm at INRIM is given \cite{Boella91} by the dc quantum Hall effect on the $i=2$ step, $R_\mathrm{K}/2 \approx $ \SI{12906.4}{\ohm}. A dc potentiometer \cite{Marullo87} performs calibrations of resistance standards. A time series of measurements of $R_\mathrm{Q}$ is shown in Fig. \ref{fig:QuadrifilarDrift}: a significant, but predictable, drift of \SI{5}{n\ohm \, \ohm^{-1} d^{-1}} is estimated.

\subsection{Characterization of the polyphase generator}

As shown in Sec. \ref{sec:RatioBridge} and \ref{sec:QuadBridge}, the reading of each bridge is given by a mathematical expression whose input quantity is the complex ratio of the nominal settings of two generators ($V_\mathrm{c}/V_\mathrm{a}$ for the ratio bridge, $V_\mathrm{c}/V_\mathrm{Q}$ for the quadrature bridge). The tracking of the different outputs of the generator (under proper loading conditions) has been adjusted and calibrated; the deviations from nominal values are within a few parts in $10^4$. Since the impedance standards deviate from their nominal values by less than a few parts in $10^{-5}$, and their relative phases have been adjusted, the contribution to final accuracy of each bridge caused by the generators can be kept near \SI{1E-8}{}. 

\begin{figure}[ht]
    \centering
    \includegraphics[width=3.5 in]{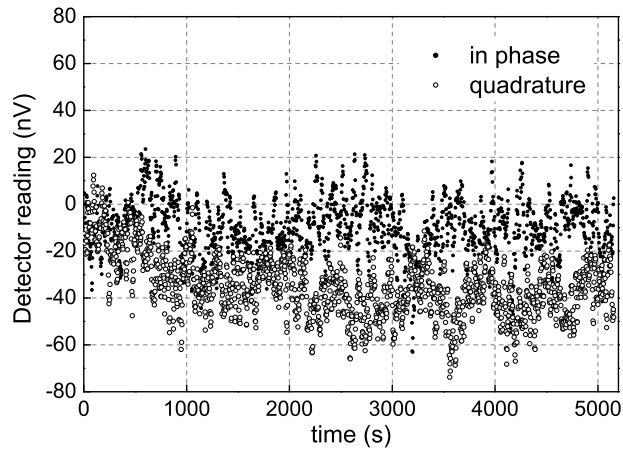}
	\caption{Typical time evolution of the detector reading (in-phase and quadrature components) at the combining network detection point of the  quadrature bridge, after an equilibrium operation (for $t=0$).}
    \label{fig:DetectorNoise}
\end{figure}

The stability and noise of the polyphase generator can be inferred from drifts of detector readings of the bridge after an equilibrium. Fig. \ref{fig:DetectorNoise} show the time evolution of the detector reading at the combining network of the quadrature bridge, which is affected by all generator output drifts.

\subsection{Ratio bridge measurements}
\label{sec:RatioMeasurements}

\begin{figure}[ht]
    \centering
    \includegraphics[width=3.5 in]{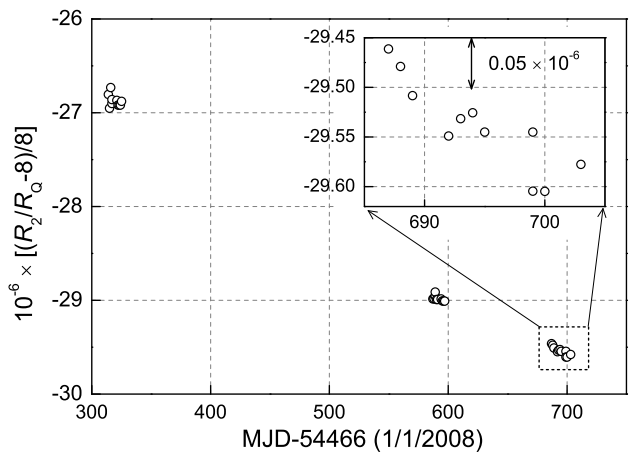}
    \caption{Results of measurement of $R_2/R_\mathrm{Q}$ with the 8:1 resistance ratio bridge over a period of 400 days. Data is expressed as relative deviation (in parts per $10^6$) from the nominal ratio $[R_2/R_\mathrm{Q}]_\mathrm{nominal} = 8$.}
    \label{fig:RatioMeasurements}
\end{figure}

Fig. \ref{fig:RatioMeasurements} shows the measurement of $R_2/R_\mathrm{Q}$ with the resistance ratio bridge (the result of $R_1/R_\mathrm{Q}$, not shown, is very similar) over a period of more than one year of operation. The ratio drift is caused by the compound drift of both $R_2$ and $R_\mathrm{Q}$. The inset of Fig. \ref{fig:RatioMeasurements} shows the repeatability of measurements over a few days. 

\subsection{Quadrature bridge measurements}
\label{sec:QuadMeasurements}

\begin{figure}[ht]
    \centering
    \includegraphics[width=3.5 in]{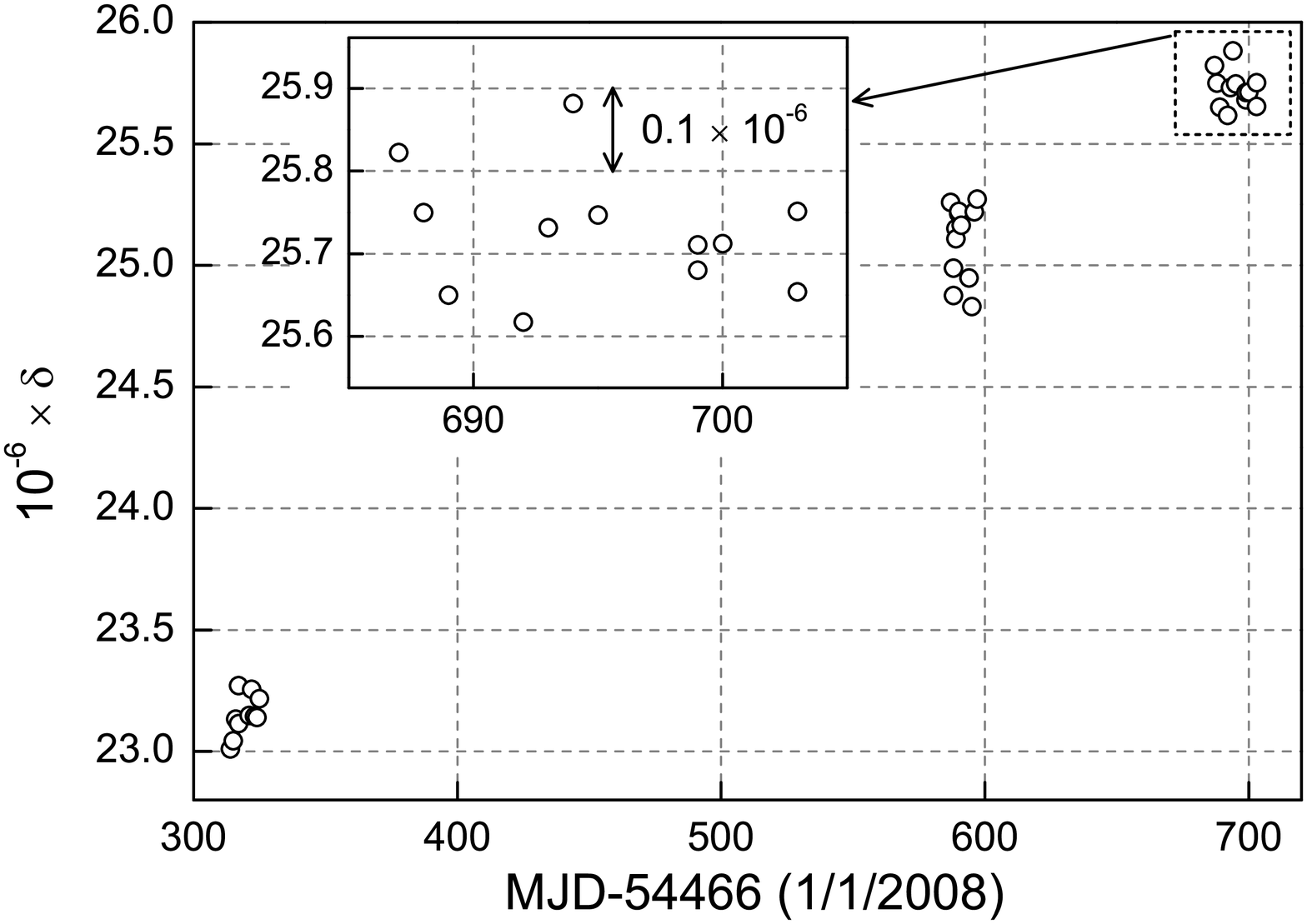}
    \caption{Results of measurement of $\delta \equiv \omega^2R_1R_2C_1C_2 - 1$ with the quadrature bridge over a period of 400 days. The inset shows the repeatability of the measurement over a few days.}
    \label{fig:QuadMeasurements}
\end{figure}

Fig. \ref{fig:QuadMeasurements} shows the measurement of $\delta$ (see Eq. \ref{quadbridgeequilibrium}) with the quadrature bridge over the same time period of Fig. \ref{fig:RatioMeasurements}. The drift is the compound drift of the standards $R_1$, $R_2$, $C_1$, $C_2$. The inset of Fig. \ref{fig:RatioMeasurements} shows the repeatability of measurements over a few days. 

\subsection{Final result, and comparison with the maintained capacitance standard}
\label{sec:comparison}

\begin{figure}[ht]
    \centering
    \includegraphics[width=3.5 in]{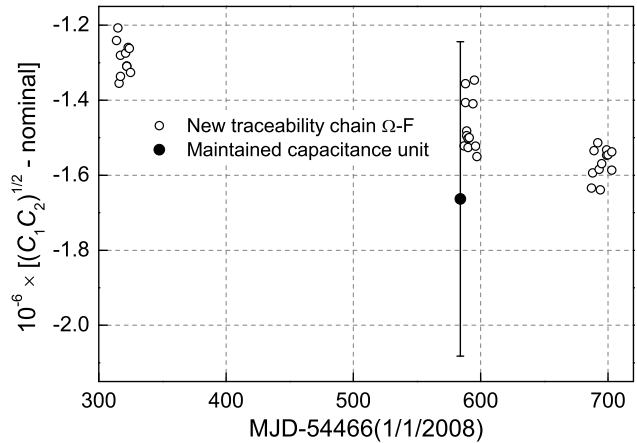}
    \caption{Comparison between quadrature bridge and step up procedure.}
    \label{fig:Comparison}
\end{figure}

Fig. \ref{fig:Comparison} shows the geometric mean of $C_1$ and $C_2$ from the nominal (\SI{1000}{pF}) value. The result is computed from all data previously described. The observed drift is attributed to coumpond drift of $C_1$ and $C_2$. 

In the same figure, the result (with uncertainty bars) given by a step-up measurement from the maintained national capacitance standard is shown; a visual agreement can be appreciated. This measurement is performed at the frequency of \SI{1592}{Hz} and should be corrected to \SI{1541}{Hz} because of the frequency dependence of $C_1$ and $C_2$. Indirect measurements of such dependence, performed with the so-called \emph{S-matrix} method \cite{Callegaro2003HF} give a correction below \SI{1E-9}{}. An uncertainty contribution associated with the correction has been nevertheless added to the uncertainty budget (see Sec. \ref{sec:uncertainty}).

\section{Uncertainty}
\label{sec:uncertainty}

Tables \ref{tab:UncSteps}--\ref{tab:comparison} give the uncertainty expression corresponding to the measurements described in Sec. \ref{sec:results}:
\begin{itemize}
\item Tab. \ref{tab:UncSteps} lists the uncertainty contributions related to the various measurements and standards employed in the new traceability chain;
\item Tab. \ref{tab:uncert_summary} gives the uncertainty budget for the measurement of the geometric mean $(C_\mathrm{1}C_\mathrm{2})^{1/2}$ of the \SI{1000}{pF} capacitance standards $C_\mathrm{1}$ and $C_\mathrm{2}$ in terms of the INRIM representation of the ohm given by the quantum Hall effect in dc regime;
\item Tab. \ref{tab:comparison} gives the uncertainty budget for the comparison described in Sec. \ref{sec:comparison} and shown in graphic form in Fig. \ref{fig:Comparison}.
\end{itemize}

\begin{table}[h]
	\caption{Measurement steps of the ohm-farad traceability chain: relative uncertainty expression (contributions and root-sum-square RSS)}
	\centering
	\begin{tabular}{l|c|c|l}
		\hline \hline
		Source of uncertainty & Type & $u_\mathrm{r}$ &Note \\
		\hline
		\textbf{1: Calibration of $\textit{R}_Q$ @ \SI{1541}{Hz}} &  & \SI{}{n\Omega \cdot \Omega^{-1}} &  \\
		\small DC calibration of $R_\mathrm{Q}$ & A, B & 25 &\footnotesize Calibration with dc QHE  \\
		\small phase correction of $R_\mathrm{Q}$ & B & \SI{12}{}  &\footnotesize \SI{1E-5}{} loss angle of \SI{10}{pF} capacitor \\
		\small Frequency dependence & B & 3  &\footnotesize  10\%  of calculated frequency deviation   \\
		\small Short-term stability  & B & 10  &\footnotesize Estimated drift is $5 \times10^{-9}$/day  \\
		\hline
		\small\textbf{RSS}& &\SI{30}{} & \\

		\hline
		\hline
		
		\textbf{2: Resistance ratio bridge} & & \SI{}{n\Omega \cdot \Omega^{-1}} & \\

		\small Noise & A & 10 &\footnotesize Std of the mean of 10 measurements  \\
		\small Main balance injection & B & 12 & \footnotesize \SI{5E-4}{} of a \SI{25E-6}{} injection \\
		\small 4TP impedance definition & B & 10 &   \\
		\small 4TP cable corrections & B & 1 &\footnotesize  See \cite{Kibble84}  \\
		\small 2TP contact resistance repeatability & B & 1 &\footnotesize BPO repeatability, \SI{100}{\micro \Omega} \\
		\small Residual loading on main IVD & B & 10 &\footnotesize  \\ 
		\small Main IVD ratio & B & 45 &\footnotesize Bootstrap calibration \cite{Callegaro2003IVD} \\
		\small Noncoaxiality & B & 5 &\footnotesize  \\
		
		\hline
		\small \textbf{RSS}& & 50 & \footnotesize{Calibration of $R_\mathrm{1}$ and  $R_\mathrm{2}$} \\

		\hline\hline
		\textbf{3: Quadrature bridge} & &  \SI{}{nF \cdot F^{-1}} & \\

		\small Noise & A & 20 &\footnotesize Std of the mean of 10 measurements  \\
		\small Frequency & B & \SI{0}{} &\footnotesize Lock to INRIM \SI{10}{MHz} frequency standard   \\
		\small Main balance injection & B & 12 & \footnotesize \SI{5E-4}{} of a \SI{25E-6}{} injection   \\
		\small Distortion & B & 6 &\footnotesize Harmonic amplitude and lock-in rejection ratio  \\
		\small Residual offset after inversion & B & 3 &\footnotesize Average of difference of direct and reverse meas. \\
		\small 2TP contact resistance repeatability & B & 1 &\footnotesize BPO repeatability, \SI{100}{\micro \Omega} \\
		\small 2TP capacitance repeatability & B & 1 & \footnotesize  \\
		\small Noncoaxiality & B & 5 & \footnotesize  \\

		\hline
		\small \textbf{RSS}& & 25 & \footnotesize{Calibration of $(C_\mathrm{1}C_\mathrm{2})^{1/2}$} \\
		\hline

	\end{tabular}
	\label{tab:UncSteps}
\end{table}

\begin{table}[h]
	\caption{Measurement of the geometric mean $(C_\mathrm{1}C_\mathrm{2})^{1/2}$ of the capacitance standards from dc quantum Hall effect with the new traceability chain: uncertainty expression.}

	\centering
	\begin{tabular}{l|c|l}
		\hline \hline
		Source of uncertainty & $u_\mathrm{r} \times 10^{-9}$ & Note \\
		\hline

		\small Calibration of $R_\mathrm{Q}$ @ \SI{1541}{Hz} & 25 &\footnotesize Tab. \ref{tab:UncSteps}, \#1.  \\
		\small Resistance ratio bridge &  50  &\footnotesize  Tab. \ref{tab:UncSteps}, \#2. \\
		\small Short-term stability of $R_\mathrm{1}$ and $R_\mathrm{2}$  & 10  &\footnotesize TC of \SI{2E-6}{K^{-1}}; \SI{5}{mK} std over \SI{1}{h}  \\
		\small Quadrature bridge  & 25  &\footnotesize Tab. \ref{tab:UncSteps}, \#3. \\ 
		\hline
		\small\textbf{RSS}& 64 & \\
		\hline

	\end{tabular}
	\label{tab:uncert_summary}
\end{table}

\begin{table}[h]
	\caption{Comparison between the measurement with the new traceability chain, and the maintained capacitance national standard.}

	\centering
	\begin{tabular}{l|l|c|l}
		\hline \hline
		Source of uncertainty & Type & $u_\mathrm{r}$ &Note \\
		\hline
		\textbf{1: Calibration of $\textit{R}_Q$ @ \SI{1541}{Hz}} &  & \SI{}{nF \cdot F^{-1}} &  \\

		\small \SI{10}{pF} maintained group value at \SI{1592}{Hz} & B & 400 &\footnotesize  \\
		\small Capacitance bridge, \SI{10}{pF} to \SI{100}{pF} step-up & B & 40  &\footnotesize \\
		\small Capacitance bridge, \SI{100}{pF} to \SI{1000}{pF} step-up  & B & 100  &\footnotesize 2 measurements  \\
		\small Short-term stability of \SI{1000}{pF} capacitors  & B & 4  &\footnotesize TC \SI{4E-6}{K^{-1}}; \SI{1}{mK} controller stability \\
		\small Frequency correction  & B & 5 &\footnotesize \SI{1541}{Hz} to \SI{1592}{Hz}, gas-dielectric \\
		\small New traceability chain, $(C_\mathrm{1}C_\mathrm{2})^{1/2}$ & B & 64 &\footnotesize Tab. \ref{tab:uncert_summary} \\ 
		\hline
		\small\textbf{RSS}& & 419 & \\
		\hline
	\end{tabular}

	\label{tab:comparison}
\end{table}

\section{Conclusions}

The paper described a new traceability chain for the realization of the farad from the quantum Hall effect, which include two bridges, a resistance ratio bridge and a quadrature bridge, based on a polyphase sinewave generator. The bridges do not contain variable passive components like multi-decadic inductive voltage dividers or impedance decadic boxes; the equilibrium is obtained by direct digital synthesis of the necessary signals. In the present implementation the bridge operation is semi-automated and the equilibrium is reached in short time.

The total relative uncertainty of the traceability chain is estimated to be \SI{64E-9}{} at the level of \SI{1000}{pF}, therefore adequate for a national metrology institute. A first verification of the realization accuracy is given by a comparison with the maintained capacitance national standard, but an international comparison is being planned in the next months.

Future improvements of the implementation will include the installation of individually thermostatted resistance standards for $R_1$ and $R_2$, and the complete automation of the bridges with a remotely-controlled coaxial switch. Since the digital assistance of primary impedance bridges has proved as a successful approach, the realization of a digitally-assisted 10:1 ratio bridge for scaling \SI{1000}{pF} to maintained \SI{10}{pF} standards is under consideration.

\section*{Acknowledgments}

The authors are indebted with their colleagues F. Francone and D. Serazio for the physical construction of the electromagnetic devices; and to  C. Cassiago for the calibration of $R_\mathrm{Q}$.

\section*{References}
\bibliographystyle{unsrt}
\bibliography{Metrologia}  

\end{document}